\def\leti{Lense--Thirring}

\def\rfr#1{Eq.(\ref{#1})}
\def\rfrs#1#2{Eqs.(\ref{#1})-(\ref{#2})}
\def\leti{Lense--Thirring}

\def\eqi{\begin{equation}}
\def\eqf{\end{equation}}
\def\eqia{\begin{eqnarray}}
\def\eqfa{\end{eqnarray}}

\def\rp#1#2{{#1\over#2}}

\def\lb#1{\label{#1}}

\def\kepa{\dot\Psi^{\rm A}_{.{\ell}}}
\def\kepaa{\dot\Psi^{\rm A}_{.{\ell^{'}}}}
\def\kepb{\dot\Psi^{\rm B}_{.{\ell}}}
\def\kepbb{\dot\Psi^{\rm B}_{.{\ell^{'}}}}
\def\kepc{\dot\Psi^{\rm C}_{.{\ell}}}
\def\kepcc{\dot\Psi^{\rm C}_{.{\ell^{'}}}}
\def\kepalt{\dot\Psi^{\rm A}_{\rm GTR}}
\def\kepblt{\dot\Psi^{\rm B}_{\rm GTR}}
\def\kepclt{\dot\Psi^{\rm C}_{\rm GTR}}

\documentclass[11pt]{article}
\usepackage{amsmath,amsthm,amscd,amssymb}
\usepackage{latexsym}
\usepackage{graphics,graphicx}
\usepackage{setspace}

\begin{document}

\noindent{\bf \LARGE{A new approach for investigating secular
variations of the low-degree geopotential coefficients}}
\\
\\
\\
{Lorenzo Iorio}\\
{\it Viale Unit\`{a} di Italia 68, 70125\\Bari, Italy
\\Email: lorenzo.iorio@libero.it\\
Phone/Fax: ++39 080 5443144}

\begin{abstract}
Recent tests aimed at the detection of the general relativistic
gravitomagnetic Lense-Thirring effect in the gravitational field
of the Earth with the LAGEOS and LAGEOS II satellites have proved
to be affected, among other things, by the corrupting impact of
the secular variations $\dot J_4$ and $\dot J_6$ of the second and
third even zonal harmonic coefficients of the multipolar expansion
of the terrestrial gravitational potential. Unfortunately, they
are, at present, poorly known so that their impact on the
performed measurement is $\sim 13\%$ over 11 years. Also the
impact of the static part of $J_4$ and $J_6$ is relevant inducing
a systematic bias of 4-9$\%$ according to different  gravity
models. Moreover, the currently available values for them may
retain an a priori `memory' effect of the Lense-Thirring effect
itself. In this paper we suggest a novel method to determine
$J_2,J_4,J_6$ along with their secular variations $\dot J_{2},\dot
J_4,\dot J_6 $. Our approach is based on the use of three linear
combinations of the nodes of LAGEOS, LAGEOS II and Ajisai and the
perigee of LAGEOS II. The retrieved values for the even zonal
harmonics of interest are, by construction, independent of each
other and of the post-Newtonian precessions. The so obtained
mini-model could subsequently be used in order to enhance the
accuracy and the reliability of the measurements of the
Lense-Thirring effect by means of LAGEOS and LAGEOS II satellites.
Moreover, it would also allow for a clearer investigation of
possible seasonal and interannual variations of $J_4$ and $J_6$.
\end{abstract}

\newpage

\section{Introduction}
The long and accurate records of data available from the
laser-ranged geodetic satellites  have proven to be an excellent
and unique tool for investigating the long-term variations of the
terrestrial gravity field by means of the satellite laser ranging
(SLR) technique. For example, Cheng and Tapley (2004) recently
analyzed 28 years of data from 1976 to 2003.

In particular, the secular variations $\dot J_{\ell}$ of the even
zonal harmonic coefficients $J_{\ell}$ of the Newtonian multipolar
expansion of the Earth's gravitational potential have recently
received attention (Cheng et al. 1997; Bianco et al. 1998; Cox and
Chao 2002, Dickey et al. 2002, Cox et al. 2003, Cheng and Tapley,
2004). This is mainly due to the observed inversion of the rate of
change of the Earth's quadrupole mass moment coefficient $J_2$
which, since 1998, began increasing (Cox and Chao 2002).  It is
not yet clear if such an effect is a long-term feature or is
short-term in nature. The variations of the even zonal harmonics
are mainly related to the Earth's lower mantle viscosity features
(Ivins et al. 1993). Recent studies have shown that also ice mass
losses, and oceanographic and hydrological contributions are
important.

An accurate knowledge of the secular variations of the other
low-degree even zonal harmonics, with particular emphasis on $\dot
J_4$ and $\dot J_6$, is of the utmost importance for a reliable
and consistent evaluation of the total accuracy in the tests aimed
at the detection of the post-Newtonian  gravitomagnetic
\leti\ effect (Lense and Thirring 1918). It consists of secular
precessions affecting the longitude of the ascending node $\Omega$
and the argument of perigee $\omega$ of the orbit of a test
particle moving in  the gravitational field of a central  spinning
body. They are\eqi\dot\Omega_{\rm LT}=\rp{2GS}{c^2
a^3(1-e^2)^{3/2}},\ \dot\omega_{\rm LT}=-\rp{6GS\cos i}{c^2
a^3(1-e^2)^{3/2}},\eqf in which $G$ and $c$ are the Newtonian
constant of gravitation and the speed of light in vacuum,
respectively, $S$ is the proper angular momentum of the central
body, $a,e,i$ are the semimajor axis, the eccentricity and the
inclination to the body's equator, respectively, of the test
particle's orbit. For the LAGEOS satellites, whose orbital
parameters are in Table \ref{orpar}, they amount to a few tens of
milliarcseconds per year (mas yr$^{-1}$).  The major source of
bias is represented by the corrupting effects induced by the
Newtonian part of the terrestrial gravitational field. A strategy
to overcome this problem was put forth for the first time by
Ciufolini (1996) who designed a linear combination involving the
nodes of LAGEOS and LAGEOS II and the perigee of LAGEOS II in
order to remove the bias due to the first two even zonal harmonics
$J_2$ and $J_4$. In the tests performed with such a combination
(Ciufolini et al. 1998) and the Earth's gravity model EGM96
(Lemoine et al. 1998) a total error of 20$\%$ was claimed.
However, this estimate is largely optimistic mainly due to the
impact of the non-gravitational perturbations affecting the
perigee of LAGEOS II and of the uncancelled even zonal harmonics.
More realistic evaluations point toward $\sim50-80\%$ (Ries et al
2003a; Iorio 2005a). Ries et al. (2003b) suggested for the first
time to use only the nodes of LAGEOS and LAGEOS II for measuring
the Lense-Thirring effect in view of the expected improvements in
our knowledge of the Earth's gravitational field from the GRACE
missions. In (Iorio 2003a; Iorio and Morea 2004; 2005c) the
following combination of the residuals of the nodes of LAGEOS and
LAGEOS II was explicitly proposed \eqi\delta\dot\Omega^{\rm
L}+p_1\delta\dot\Omega^{\rm L\ II }, \lb{iorform}\eqf with
$p_1=0.546$ (see Section \ref{combilin}). \rfr{iorform} would
entirely absorb the gravitomagnetic signature, which is a linear
trend with a slope of 48.1 mas yr$^{-1}$, because the
Lense-Thirring effect is purposely switched off in the dynamical
force models used for constructing the (O)-(C) residuals
$\delta\dot\Omega$. The coefficient $p_1$ allows to cancel out the
aliasing effects due to the static and time-varying components of
the first even zonal harmonic $J_2$. The combination of
\rfr{iorform} is, instead, affected by the other higher degree
even zonal harmonics $J_4, J_6, J_8..$ along with their secular
variations $\dot J_4, \dot J_6$. While the static parts of
$J_{\ell}$ induce linear precessions, the shift due to the secular
variations $\dot J_{\ell}$ is quadratic. Ciufolini and Pavlis
(2004) adopted the combination of \rfr{iorform} for analyzing
almost 11 years of LAGEOS and LAGEOS II data with the 2nd
generation GRACE-only EIGEN-GRACE02S model (Reigber et al. 2005).
They claim a total error of 5$\%$ at 1-sigma level. Such estimates
have been criticized by Iorio (2005a; 2005b) who yields a
19-24$\%$ 1-sigma total error budget estimate. In (Iorio 2005b) it
has been shown that the systematic error due to the static part of
the even zonal harmonics is still model-dependent ranging from
4.3$\%$ for the GeoForschungsZentrum (GFZ) EIGEN-GRACE02S model to
8.7$\%$ for the Center for Space Research (CSR) GGM02S model (see
http://www.csr.utexas.edu/grace/gravity/). These figures refer to
1-sigma upper bounds obtained by linearly summing the individual
mismodelled precessions. It is important to note that the
sensitivity of GRACE to the low-degree even zonal harmonics has
recently been questioned (Wahr et al. 2004); this could prevent
from obtaining notable future improvements of the systematic error
in the Lense-Thirring measurement with \rfr{iorform} due to the
static part of the even zonals.

In regard to the secular variations of the even zonal harmonics,
at present there is a large uncertainty about the magnitude and
even the sign of $\dot J_4$ and $\dot J_6$; see (Cox et al. 2003,
Table 1). Thus, their impact over the performed measurement has
been evaluated 13$\%$ over 11 years (Iorio 2005b).

In this paper, we propose a strategy in order to measure the first
three even zonal harmonics of the geopotential independently of
each other and of the relativistic effects by means of suitable
linear combinations of the orbital residuals of the LAGEOS
satellites and of Ajisai. This approach could also be useful in
better analyzing the temporal features of $J_4$ and $J_6$
independently of $J_2$.
\section{The linear combination approach and its main
features}\lb{combilin} To determine the even zonal harmonic
$J_{\ell}$, it is convenient to consider the temporal evolution,
averaged over many orbital revolutions, of those Keplerian orbital
elements which, under such condition, undergo secular precessions.
They are the node $\Omega$, the longitude of the perigee $\omega$
and the mean anomaly $\mathcal{M}$; for a generic satellite A we
will denote any of them as $\Psi^{\rm A}$. Among them, the node
$\Omega$ is by far the most accurately measurable, mainly due to
its insensitivity to the non-gravitational perturbations. On the
contrary, the perigee $\omega$ is affected by a host of
non--gravitational perturbations which, in many cases, are
difficult to correctly model so that their impact on the recovery
of some parameter of interest by means of the perigee cannot often
be reliably assessed and evaluated (Ries et al 2003a; 2003b). The
same holds also for the mean anomaly $\mathcal{M}$ for which the
indirect perturbations on the mean motion $n$ induced by the
disturbing accelerations affecting the semimajor axis $a$ are to
be considered as well.

It is well known that the multipolar expansion of the Earth's
geopotential in spherical harmonics (Kaula 1966) leads to
classical secular precessions of the node, the perigee and the
mean anomaly which are linear in the even zonal harmonics
$J_{\ell}$. Moreover, the Einstein's general theory of relativity
(GTR) predicts that additional secular precessions affecting the
node and/or the perigee are also present. They are
 the gravitoelectric Einstein secular rate of the perigee (Einstein
1915) \eqi\dot\omega_{\rm GE}=\rp{3nGM}{c^2 a(1-e^2)},\eqf in
which $M$ is the central body's mass and $n=\sqrt{GM/a^3}$ is the
Keplerian mean motion, and the gravitomagnetic Lense-Thirring
rates of the node and the perigee. In Table \ref{lent} we quote
the post-Newtonian secular precessions for the nodes of LAGEOS,
LAGEOS II and Ajisai and the perigee of LAGEOS II.

Let us suppose, for the sake of clarity, that we want to measure
two particular even zonal harmonics $J_{\ell}$ and $J_{\ell^{'}}$
in a relativity--free fashion. The orbital residuals of a given
Keplerian element account for any mismodelled or unmodelled
physical phenomenon affecting that element. In order to illustrate
the method, let us consider the orbital residuals
$\delta\dot\Psi_{\rm obs}$ of the rates $\dot\Psi$ for three
satellites denoted as A, B and C and assume that they entirely
account for one of the three features that we wish to measure
separately of each other, (i.e., for the relativistic effects, or
the even zonal harmonic of degree ${\ell}$ or ${\ell^{'}}$).
In regard to the even zonal harmonics, this means that we are
using a  `truncated' Earth gravity model in the force model
routines of the orbit determination system; in principle, it would
include all the even zonal harmonics except for that one in which
we are interested in.
We can write \eqi \left\{
\begin{array}{lll}
\delta\dot\Psi^{\rm A}_{\rm obs}&=&\kepa J_{\ell}+\kepaa
J_{\ell^{'}}+\kepalt\mu_{\rm GTR}+\Delta^{\rm A},\lb{uno}\\
\delta\dot\Psi^{\rm B}_{\rm obs}&=&\kepb J_{\ell}+\kepbb
J_{\ell^{'}}+\kepblt\mu_{\rm GTR}+\Delta^{\rm B},\\
\delta\dot\Psi^{\rm C}_{\rm obs}&=&\kepc J_{\ell}+\kepcc
J_{\ell^{'}}+\kepclt\mu_{\rm GTR}+\Delta^{\rm C},
\end{array}
\right. \eqf
 in which the coefficients $\dot\Psi_{.{\ell}}$ are
defined as \eqi\dot\Psi_{.{\ell}}=\frac{\partial \dot\Psi_{\rm
class}}{\partial J_{{\ell}}}.\eqf The coefficients
$\dot\Psi_{.{\ell}}$ have been explicitly worked out for
$\Psi\equiv\Omega$ and $\Psi\equiv\omega$ from $\ell=2$ to
$\ell=20$ (Iorio 2003b); it turns out that they are functions of
the semimajor axis $a$, the inclination $i$ and the eccentricity
$e$ of the considered satellite:
$\dot\Psi_{.{\ell}}=\dot\Psi_{.{\ell}}(a,\ e,\ i;\ GM,R)$, where
$R$ is the Earth's mean equatorial radius. For example, the
coefficient of degree $\ell=2$ for the node is
\eqi\dot\Omega_{.2}=-\rp{3}{2}\rp{n\cos i
}{(1-e^2)^2}\left(\rp{R}{a}\right)^2,\eqf while for the perigee we
have
\eqi\dot\omega_{.2}=\rp{3}{4}\rp{n}{(1-e^2)^2}\left(\rp{R}{a}\right)^2\left(5\cos^2
i-1 \right).\eqf In Table \ref{param} we quote the coefficients
$\dot\Psi_{.\ell}$ for the nodes of LAGEOS, LAGEOS II and Ajisai
and the perigee of LAGEOS II from $\ell=2$ to $\ell=6$.

The quantities $\Delta$ include all the other classical effects,
of gravitational and non--gravitational origin, which affect
$\Psi$ and which have been included in the routines of the orbit
determination systems  like, e.g., GEODYN II (Pavlis et al. 1998)
or UTOPIA (CSR) with the level of accuracy (or mismodelling, if
you prefer) characteristic of their models. If such force models
were perfect, such other effects would not affect the residuals
$\delta\dot\Psi$ and we would have $\Delta=0$. In particular, we
can assume they include the mismodelled secular precessions
induced by the remaining even zonal harmonics of degree other than
${\ell}$ or ${\ell^{'}}$, possible seasonal and interannual
variations and the mismodelled non-gravitational perturbations.
The quantity $\mu_{\rm GTR}$ is a solved--for least square
parameter which accounts for the post--Newtonian effects. It is 0
in classical mechanics and 1 in GTR.
We can consider \rfr{uno} as a non--homogeneous algebraic linear
system of three equations in the three unknowns $J_{{\ell}}$,
$J_{{\ell^{'}}}$ and $\mu_{\rm GTR}$. The square matrix of
coefficients is represented by
\begin{equation}
\left(
\begin{array}{lll}
  \kepa & \kepaa & \kepalt \\
  \kepb & \kepbb & \kepblt \\
  \kepc & \kepcc & \kepclt
\end{array}
\right)
\end{equation}
By defining

  \eqi\mathcal{A} \equiv  \delta\dot\Psi^{\rm A}_{\rm obs}-\Delta^{\rm A},\eqf
  \eqi\mathcal{B}  \equiv  \delta\dot\Psi^{\rm B}_{\rm obs}-\Delta^{\rm B},\eqf
  \eqi\mathcal{C}  \equiv  \delta\dot\Psi^{\rm C}_{\rm obs}-\Delta^{\rm C},\eqf

it is possible to obtain
\begin{eqnarray}
  J_{\ell}&=&\frac{\mathcal{A}+h_1\mathcal{B}+h_2\mathcal{C}}{\kepa+h_1\kepb+h_2\kepc},\lb{J2}\\
  J_{\ell^{'}}&=&\frac{\mathcal{A}+k_1\mathcal{B}+k_2\mathcal{C}}{\kepaa+k_1\kepbb+k_2\kepcc}\lb{J4},\\
  \mu_{\rm
  GTR}&=&\frac{\mathcal{A}+c_1\mathcal{B}+c_2\mathcal{C}}{\kepalt+c_1\kepblt+c_2\kepclt}\lb{grslope},
\end{eqnarray} where
\begin{eqnarray}
h_1&=&\frac{\kepalt\kepcc-\kepclt\kepaa}{\kepclt\kepbb-\kepblt\kepcc},\\
h_2&=&\frac{\kepblt\kepaa-\kepalt\kepbb}{\kepclt\kepbb-\kepblt\kepcc},\\
k_1&=&\frac{\kepclt\kepa-\kepalt\kepc}{\kepblt\kepc-\kepclt\kepb},\\
k_2&=&\frac{\kepalt\kepb-\kepblt\kepa}{\kepblt\kepc-\kepclt\kepb},\\
c_1&=&\frac{\kepc\kepaa-\kepa\kepcc}{\kepb\kepcc-\kepc\kepbb},\lb{cf1}\\
c_2&=&\frac{\kepa\kepbb-\kepb\kepaa}{\kepb\kepcc-\kepc\kepbb}\lb{cf2}.
\end{eqnarray}
This approach was proposed for the first time by Ciufolini (1996)
in the context of the \leti\ experiment with the LAGEOS
satellites. He derived a combination involving the nodes of the
LAGEOS satellites and the perigee of LAGEOS II. It is a particular
case of \rfr{grslope} and \rfrs{cf1}{cf2} in which $\Psi^{\rm
A}=\Omega^{\rm LAGEOS}$, $\Psi^{\rm B}=\Omega^{\rm LAGEOS\ II}$,
$\Psi^{\rm C}=\omega^{\rm LAGEOS\ II}$, $\ell=2$, $\ell^{'}=4$,
$\dot\Omega_{\rm GTR}=\dot\Omega_{\rm LT}$, $\dot\omega_{\rm
GTR}=\dot\omega_{\rm LT}$. Later, other combinations have been
proposed (Iorio 2003a; Iorio and Morea 2004; Iorio 2005c; Iorio
and Doornbos 2005d). For example, the coefficient $p_1$ of the
combination of \rfr{iorform} is \eqi p_1=-\rp{\dot\Omega^{\rm
L}_{.2}}{\dot\Omega^{\rm L II}_{.2}}.\eqf

It is very important to note that \rfrs{J2}{J4} allow to measure
the selected even zonal harmonics $J_{\ell}$ and $J_{\ell^{'}}$
independently of each other and of the relativistic effects which
are a priori assumed to be valid and included in the force models.

\section{Three possible combinations for ${J_2}$, ${J_4}$ and
${J_6}$}

Applying the considerations outlined in Section \ref{combilin} to
the first three even zonal harmonics, it is possible to obtain
three linear combinations which allow to determine them
independently of each other and of the relativistic effects. Note
that  the retrieved values for $\dot J_{\ell}$ are instead
correlated in, e.g., (Cheng et al. 1997): the correlation
coefficients are -0.75 for $\dot J_2$ and $\dot J_4$, 0.76 for
$\dot J_2$ and $\dot J_6$ and -0.86 for $\dot J_4$ and $\dot J_4$
and $\dot J_6$. We will use the LAGEOS satellites and Ajisai
because of their high altitude which allow to reduce the impact of
the remaining uncancelled even zonal harmonics $J_8, J_{10},
J_{12},...$ which do affect the proposed combinations. The direct
inclusion of the other geodetic satellites, like Starlette
($a=7331$ km, $i=49.8$ deg, $e=0.0205$), which has a rather
eccentric orbit, and Stella ($a=7193$ km, $i=98.6$ deg, $e=0$)
would introduce more even zonals due to their semimajor axes and
inclinations, thus enhancing the systematic error induced by the
uncertainties in such other even zonal coefficients.

The combination for $J_2$ is \eqi\delta\dot\Omega^{\rm
L}+a_1\delta\dot\Omega^{\rm L\ II}+a_2\delta\dot\Omega^{\rm
Aji}+a_3\delta\dot\omega^{\rm L\ II},\lb{J2}\eqf with $a_1 =
2.865127386220304,a_2 = -0.1125697464440513,a_3 =
-0.03288044638241112.$ The expected signal affecting \rfr{J2} is
\eqi X\left(J_2+\dot J_2 t\right)+[{\rm seasonal,\ interannual\
effects}]+[\delta J_{8}, \delta J_{10},... ]+[{\rm
nongrav.}],\lb{X}\eqf
 with $X\equiv\dot\Omega^{\rm L}_{.2}+a_1\dot\Omega^{\rm L\ II}_{.2}+a_2\dot\Omega^{\rm Aji}_{.2}+a_3
 \dot\omega^{\rm L\ II}_{.2}=-1046.48249580121\ {\rm deg\
day}^{-1}.$ For $a_1,a_2,a_3,X$ the values in Table \ref{param}
and Table \ref{lent} have been used.
\rfr{J2} is purposely built up in order to cancel out the
classical perturbations induced by $J_4, J_6$, along with their
temporal variations, and the post-Newtonian secular precessions.
This feature can easily be checked by calculating \rfr{J2} with
the coefficients $\dot\Psi_{.\ell}$ of the classical precessions
of degree $\ell=4,6$ (Table \ref{param}) and with the general
relativistic precessions (Table \ref{lent}): the result is zero.

For $J_4$ we have \eqi\dot\Omega^{\rm L }+b_1\delta\dot\Omega^{\rm
L\ II}+b_2\delta\dot\Omega^{\rm Aji}+b_3\delta\dot\omega^{\rm L\
II},\lb{J4}\eqf with $b_1 =
0.7547006806958291,b_2=-0.04496658671744361,b_3=-0.01501043654953821.$
The expected signal for \rfr{J4} is \eqi Y\left(J_4+\dot J_4
t\right)+[{\rm seasonal,\ interannual\ effects}]+[\delta J_{8},
\delta J_{10},... ]+[{\rm nongrav.}],\lb{Y}\eqf with $Y\equiv
\dot\Omega^{\rm L}_{.4}+b_1\dot\Omega^{\rm L\
II}_{.4}+b_2\dot\Omega^{\rm Aji}_{.4}+b_3
 \dot\omega^{\rm L\ II}_{.4}=-86.5180659848\ {\rm deg\
day}^{-1}$. For $b_1,b_2,b_3,Y$ the values in Table \ref{param}
and Table \ref{lent} have been used. \rfr{J4} is, by construction,
independent of $J_2$, $J_6$ and relativity.

The even zonal harmonic coefficient $J_6$ can be determined
independently of $J_2$, $J_4$ and relativity by means of
\eqi\delta\dot\Omega^{\rm L}+c_1\delta\dot\Omega^{\rm L\
II}+c_2\delta\dot\Omega^{\rm Aji}+c_3\delta\dot\omega^{\rm L\
II},\lb{J6}\eqf with $c_1 = 5.747950720464317,
c_2=-1.073942838094944, c_3=-0.02646944285900744.$ It is expected
that \rfr{J6} is affected by \eqi Z\left(J_6+\dot J_6
t\right)+[{\rm seasonal,\ interannual\ effects}]+[\delta J_{8},
\delta J_{10},... ]+[{\rm nongrav.}],\lb{Z}\eqf with $Z\equiv
\dot\Omega^{\rm L}_{.6}+c_1\dot\Omega^{\rm L\
II}_{.6}+c_2\dot\Omega^{\rm Aji}_{.6}+c_3
 \dot\omega^{\rm L\ II}_{.6}=-1022.9930383702\ {\rm deg\
day}^{-1}$. Also in this case  the values in Table \ref{param} and
Table \ref{lent} have been used for calculating $c_1,c_2,c_3,Z$.
\section{The analysis strategy} The residuals of the nodes of LAGEOS, LAGEOS II and Ajisai and of the perigee of
LAGEOS II should be built up by adopting, as usual in precise
orbit determination process, a complete suite of dynamical force
models and some background reference model for the Earth's
gravitational field complete to all available degree and order,
apart from just the even zonal harmonic coefficients that we are
interested in. Alternatively, it would be possible to include in
the reference models default values $J_{\ell}^{(0)}$ for the even
zonals of interest in order to determine corrections $\Delta
J_{\ell}$ to them. In this way, the residuals should entirely-or
partly-account for the linear and quadratic shifts induced by that
even zonal harmonic, which we are interested in, and for the all
other physical effects, which we are not interested in, according
to the level of accuracy of the force models included in the orbit
determination system. For the background reference model of the
gravitational field it would be better to use some Earth gravity
models derived from non-SLR datasets like records from CHAMP and
GRACE.

The so built residuals time series should be analyzed over a time
span rather long to average out the various time-dependent
perturbations, in particular those induced by the
non-gravitational forces acting on the perigee of LAGEOS II. In
addition to the expected linear and quadratic integrated shifts of
\rfr{X}, \rfr{Y} and \rfr{Z} also seasonal and interannual
variations could affect the investigated signals. In (Cheng and
Tapley 2004) it has been shown that such additional features
mainly affect $J_2$: our strategy could allow to better
investigate their effects on $J_4$ and $J_6$ which should be less
relevant.

With a quadratic fit $QF=a_0+b_0t+c_0t^2$ it is possible to
retrieve both $J_{\ell}$ (or $\Delta J_{\ell}$) and $\dot
J_{\ell}$ for the considered even zonal harmonic from $b_0$ and
$c_0$.
\subsection{Possible sources of errors in $J_2, J_4, J_6$}
The accuracy of the orbit determination process sets the level of
the obtainable observational errors in the even zonal harmonics
$J_{\ell}^{(\rm obs)}$. The rms orbital accuracy $delta r$ for the
LAGEOS satellites is $\lesssim 1$ cm; for Ajisai, which is more
sensitive to the non-gravitational perturbations, we
conservatively assume 10 cm. By assuming $\delta\Omega\sim\delta
r/a$ and $\delta\omega\sim\delta r/ea$, we obtain $\delta
J_2^{(\rm obs)}=2\times 10^{-14}, \delta J_4^{(\rm obs)}=9\times
10^{-14}, \delta J_6^{(\rm obs)}=2\times 10^{-12}$. over a time
span of, say, one year.

A major source of systematic error is represented by the impact of
the mismodelled even zonal harmonics of higher degree which are
not cancelled out by \rfrs{J2}{J6}. For example, \rfr{J4} is
independent of $J_2$, $J_6$ and relativity but is affected by
$J_8, J_{10}, J_{12},...$ which are assumed to be modelled
according to some Earth gravity model solution. Such bias can be
evaluated, in a conservative way, by using the available sigma
$\delta J_{\ell}$ of the variance matrix of the adopted background
reference model and by summing up the individual secular
precessions induced by the uncancelled mismodelled even zonal
coefficients. For example, the systematic error in the measurement
of $J_4$ with \rfr{J4} can be evaluated as \eqi\delta J^{(\rm
zon)}_4=\rp{\sum_{\ell=8}^{20}\left|\dot\Omega_{.\ell}^{\rm
L}+b_1\dot\Omega_{.\ell}^{\rm L\ II}+b_2\dot\Omega_{.\ell}^{\rm
Aji}+b_3\dot\omega_{.\ell}^{\rm L\ II}\right|\delta J_{\ell}
}{Y}.\lb{zon}\eqf The impact of the uncancelled even zonal
harmonics on the measurement of $\dot J_4$ with \rfr{J4} can
analogously be evaluated as \eqi\delta \dot J^{(\rm
zon)}_4=2\rp{\sum_{\ell=8}^{20}\left|\dot\Omega_{.\ell}^{\rm
L}+b_1\dot\Omega_{.\ell}^{\rm L\ II}+b_2\dot\Omega_{.\ell}^{\rm
Aji}+b_3\dot\omega_{.\ell}^{\rm L\ II}\right|\delta J_{\ell}
}{YT_{\rm obs}},\lb{zondot}\eqf where $T_{\rm obs}$ is the adopted
observational time span in years. The factor 2 in \rfr{zondot}
comes from the fact that we have compared the integrated linear
shift due to $J_8, J_{10},...$ to the quadratic shift
$Y\left(\rp{\dot J_4}{2} t^2\right)$ over $T_{\rm obs}$.

According to the combined CHAMP+GRACE+terrestrial
gravimetry/altimetry EIGEN-CG01C Earth gravity model (Reigber et
al. 2004), the systematic errors due to the uncancelled even zonal
harmonics would be $\delta J_2^{(\rm zon)}=6\times 10^{-13},
\delta J_4^{(\rm zon)}=3\times 10^{-12}, \delta J_6^{(\rm
zon)}=8\times 10^{-12}$. For the secular variations of the even
zonal harmonics of interest we have $\delta \dot J_2^{(\rm
zon)}=1\times 10^{-12}/T_{\rm obs}$ yr$^{-1}$, $\delta \dot
J_4^{(\rm zon)}=3\times 10^{-12}/T_{\rm obs}$ yr$^{-1}$, $\delta
\dot J_6^{(\rm zon)}=1\times 10^{-11}/T_{\rm obs}$ yr$^{-1}$. They
are 1-sigma upper bounds obtained, in a conservative way, by
summing up the absolute values of the individual mismodelled
precessions according to \rfr{zon} and \rfr{zondot}.

It is interesting to note that the proposed determination of $J_4$
and $J_6$, which are the most relevant with respect to the
Lense-Thirring effect determination, would not be biased by a
number of effects which, on the contrary, should be considered for
$J_2$. Among them, there is the semisecular harmonic perturbation
induced by the 18.6--year tide because its major power is
concentrated just in the $\ell=2$ $m=0$ constituent. Consequently,
it would affect, in principle, the recovery of $J_2$. According to
recent estimates of the amplitudes of the perturbations induced by
it on the nodes of the LAGEOS satellites and the perigee of LAGEOS
II and of the level of misomodelling in the $k_2$ Love number
(1.5$\%$ Iorio 2001) in terms of which they are expressed, the
18.6--year tide should have an impact on the measurement of $ J_2$
of the order of 10$^{-11}$ over one year. However, it might be a
rather pessimistic estimate. Indeed, there are sufficiently long
data records  to empirically estimate the 18.6-year tide (Eanes
and Bettadpur 1996) whose action could thus be removed from the
times series if \rfr{J2}. Indeed, the available data records of
LAGEOS and Starlette are nearly 30 years long.

In regard to other sources of systematic errors, it is important
to notice that the impact of the non--gravitational perturbations
affecting especially the perigee of LAGEOS II should be reduced by
the fact that the coefficients $a_3,b_3,c_3$ with which it enters
the three combinations are all of the order of $10^{-2}$. In order
to get an--order--of--magnitude estimate, some authors claim that
the non--gravitational perturbations affecting the perigee of
LAGEOS II could have an impact on the performed Lense--Thirring
tests of the order of 100$\%$ (Ries et al. 2003a; 2003b), i.e.
almost 60 mas yr$^{-1}$ (Ciufolini 1996). If so, the bias induced
on our proposed estimation of $ J_2,\ J_4$ and $J_6$ would be of
the order of 10$^{-13}$ over a time span of, say, one year.  In
regard to Ajisai, its node is not affected by low-frequency
perturbations which could bias the recovery of the quantities of
interest over not too long time spans. For example, the most
insidious tidal perturbation, i.e. the $\ell=2,m=1$ $K_1$ tide,
has the same period of the node which is 0.32 years for Ajisai. It
is certainly more sensitive than the LAGEOS satellites to the
non-gravitational perturbations (Sengoku et al. 1995; 1996) mainly
due to its larger area-to-mass $A/M$ ratio which is $2.7\times
10^{-2}$ m$^2$ kg$^{-1}$: for LAGEOS $A/M=6.9\times 10^{-4}$.
However, their nominal impact can be considered negligible, as can
be inferred from (Iorio and Doornbos 2005) in which it is shown
that their influence on a proposed direct measurement of the
Lense-Thirring effect is small. Moreover, note that the
coefficient $b_2$ of the combination for $J_4$, which is
particularly important for our purposes, is of the order of
$10^{-2}$.

\section{Conclusions}
Motivated by the need of improving the reliability and the
accuracy of the Lense-Thirring test performed with the node-node
LAGEOS-LAGEOS II combination of \rfr{iorform}, we have outlined a
strategy to determine $J_2, J_4, J_6 $ along with their secular
variations $\dot J_2,\dot J_4,\dot J_6$, independently of each
other and of relativity itself. Indeed, one of the major sources
of systematic error in such a relativistic test is represented by
the aliasing classical linear and quadratic shifts due to the
mismodelling in $J_4, J_6,\dot J_4,\dot J_6$. According to the
most recent CHAMP/GRACE Earth gravity models, the systematic error
induced by the static part of the zonals ranges from 4$\%$ to
$\sim 9\%$, while the impact of their secular variations is 13$\%$
over 11 years. The three linear combinations of \rfrs{J2}{J6}
built up with the nodes of LAGEOS, LAGEOS II and Ajisai and the
perigee of LAGEOS II allow to disentangle, by construction, the
relativistic precessions from those induced by the even zonals
whose measured values could thus be safely used for reanalyzing
the Lense-Thirring combination of \rfr{iorform}. In regard to
$\dot J_4$ and $\dot J_6$, it should be noted that the proposed
technique allows for a determination independent, by construction,
of the static and time-varying parts of $J_2$. Moreover, a
reanalysis of \rfr{iorform} with the values of $J_4$ and $J_6$
determined with the proposed approach would not be driven towards
the expected result by possible `imprinting' of relativistic
effects. The so obtained low-degree minimodel would also have an
intrinsic value itself, especially in view of the recently
discussed ability of GRACE to accurately measure the low-degree
even zonal harmonic coefficients (Wahr et al. 2004). Moreover, a
clearer investigation of possible seasonal and interannual
variations in $J_4$ and $J_6$ could be performed.

\newpage
\section*{Acknowledgments}
I thank L. Guerriero for his logistic support in Bari and S.
Schiller for the useful discussion in Rio at the X Marcel
Grossmann Meeting 2003. Special thanks also to the anonymous
referees due to their efforts which greatly improved the
manuscript.

\section*{Tables}
Table \ref{orpar}. Orbital parameters of LAGEOS, LAGEOS II and
Ajisai.\\
Table \ref{lent}. Post-Newtonian gravitomagnetic (LT) and
gravitoelectric (GE) secular precessions, in rad s$^{-1}$, for the
nodes of LAGEOS, LAGEOS II and Ajisai and the perigee of LAGEOS
II. The perigees of LAGEOS and Ajisai are not good observables
because of the too small eccentricities of their orbits.\\
Table \ref{param}. Newtonian secular precession coefficients
$\dot\Psi_{.\ell}=\partial\dot\Psi_{\rm class}/\partial J_{\ell}$,
in rad s$^{-1}$, for the nodes of (L), (L II) and (A) and the
perigee of LAGEOS II ($\ell=2,4,6$). The explicit expressions of
$\dot\Omega_{.\ell}$ and $\dot\omega_{.\ell}$ up to degree
$\ell=20$ can be found in Iorio (2003).
\newpage
\begin{table}
\caption{
}
\label{orpar}
\begin{tabular}{@{\hspace{0pt}}llll}
\hline\noalign{\smallskip}
 Orbital element & LAGEOS & LAGEOS II & Ajisai
\\
\noalign{\smallskip}\hline\noalign{\smallskip}
semimajor axis $a$ (km) & 12270 & 12163 & 7870\\
eccentricity $e$ & 0.0045 & 0.014 & 0.001 \\
inclination $i$ (deg) & 110& 52.65 & 50 \\
\noalign{\smallskip}\hline
\end{tabular}
\end{table}

\begin{table}
\caption{
} \label{lent}
\begin{tabular}{@{\hspace{0pt}}llll}
\hline\noalign{\smallskip}
 PN precession  & LAGEOS & LAGEOS II & Ajisai
\\
\noalign{\smallskip}\hline\noalign{\smallskip}
$\dot\Omega_{\rm LT}$ & $4.7\times 10^{-15}$& $4.8\times 10^{-15}$ & $1.79\times 10^{-14}$\\
$\dot\omega_{\rm LT}$ & - & $-8.8\times 10^{-15}$ & - \\
$\dot\omega_{\rm GE}$ & - & $5.143\times 10^{-13}$ & - \\
\noalign{\smallskip}\hline
\end{tabular}
\end{table}
\begin{table}
\caption{
} \label{param}
\begin{tabular}{@{\hspace{0pt}}llll}
\hline\noalign{\smallskip}
 $\dot\Psi_{.\ell}$ & $\ell=2$ & $\ell=4$ & $\ell=6$
\\
\noalign{\smallskip}\hline\noalign{\smallskip}
 $\dot\Omega_{.\ell}^{\rm L}$ & 6.4393531015$\times 10^{-5}$ &
2.3720267441$\times 10^{-5}$ & 4.994251263$\times
10^{-6}$\\
$\dot\Omega_{.\ell}^{\rm L\ II}$ & -1.1781974640$\times 10^{-4}$ &
-8.582111371$\times 10^{-6}$ & 7.668773045$\times
10^{-6}$\\
$\dot\Omega_{.\ell}^{\rm Aji}$ & -5.72645349511$\times 10^{-4}$ &
-2.5332565232$\times 10^{-5}$ & 2.379851554$\times 
10^{-4}$\\
$\dot\omega_{.\ell}^{\rm L\ II}$ & 8.1596225662$\times 10^{-5}$ &
6.0312610809$\times 10^{-5}$ &
 5.3628932$\times 10^{-6}$\\
\noalign{\smallskip}\hline
\end{tabular}
\end{table}

\end{document}